\documentclass[12pt]{iopart}

\usepackage{cite}
\usepackage{dsfont}
\usepackage{iopams}  
\usepackage{xcolor}
\usepackage{graphicx}
\usepackage{tikz-cd}
\usepackage{enumitem} % for customizing list environments
\usepackage{nameref}  % for \nameref
\usepackage[colorlinks=true, linkcolor=red, citecolor=blue, urlcolor=blue]{hyperref}
\hypersetup{
  pdftitle={Decoupling in Growing Populations and Feynman--Kac},
  pdfauthor={Ethan Levien, Ya\"ir Hein and Farshid Jafarpour},
  pdfproducer={pdfTeX},
  pdfdisplaydoctitle=true
}

\usepackage{tocloft}    % to fix the spacing for appendix in the Table of Contents
\usepackage{array,ragged2e} % for Table 2
\newcolumntype{P}[1]{>{\RaggedRight\arraybackslash}p{#1}} % ragged-right, keeps hyphenation; for Table 2

\expandafter\let\csname equation*\endcsname\relax
\expandafter\let\csname endequation*\endcsname\relax
\makeatother
\usepackage{amsmath}

\usepackage{orcidlink}
\newcounter{mainassump}
\newcommand{\mainassumplabel}[1]{%
  \refstepcounter{mainassump}%
  \textbf{Condition \themainassump}\label{#1}%
}

\newcounter{subassump}

\newlist{assumplist}{description}{1}
\setlist[assumplist]{labelwidth=3.5em, leftmargin=4em, align=left, labelsep=0.5em}

\newcommand{\mcL}{\mathcal L}
\newcommand{\mcA}{\mathcal A}

\newcommand{\mcM}{\mathcal M}

\newcommand{\bX}{{\boldsymbol X}}
\newcommand{\bY}{{\boldsymbol Y}}

\newcommand{\by}{{\boldsymbol y}}
\newcommand{\bx}{{\boldsymbol x}}

\newcommand{\reals}{\mathbb R}
\newcommand{\bbY}{\mathbb Y}

\newcommand{\bbP}{\mathbb P}
\newcommand{\E}{\mathbb E}
\newtheorem{theorem}{Theorem}

\begin{document}

\title[Fluctuating size and growth phenotypes]{Size-structured populations with growth fluctuations: Feynman--Kac formula and decoupling}

\author{Ethan Levien$^{1,*}$, Ya\"ir Hein$^{2,*}$ and Farshid Jafarpour$^{2,3}$\orcidlink{0000-0002-2441-0296}}% 

\address{$^1$Department of Mathematics, Dartmouth College, Hanover, NH, USA}
\address{$^2$Institute for Theoretical Physics, Department of Physics, and $^3$Centre for Complex Systems Studies, Utrecht University, Utrecht, Netherlands}

\ead{ethan.a.levien@dartmouth.edu}

\vspace{10pt}
\begin{indented}
\item[]$^*$These authors contributed equally to this work.
\end{indented}

\begin{abstract}
We study a size-structured population model in which individual cells grow at a rate determined by a fluctuating internal variable (e.g., gene expression levels). Many previous models of phenotypically heterogeneous populations can be viewed as special cases of this model, and it has previously been observed that  the internal variable decouples from cell size under certain conditions. In this work, we generalize these results and connect them to Feynman-Kac formula, which yields relationships between the lineage dynamics and population distribution in branching processes. To this end, we derive conditions for decoupling, both in the lineage and population ensemble. When decoupling occurs in both ensembles, the size dynamics can be transformed, via a random time change, into a growth-homogeneous process, and expectations can be evaluated through an exponential tilting procedure that follows from the Feynman–Kac formula. We further characterize weaker, ensemble-specific forms of decoupling that hold in either the lineage or the population ensemble, but not both. We provide a more general interpretation of tilted expectations in terms of the mass-weighted phenotype distribution. 
\end{abstract}

\maketitle

\tableofcontents

%%%%%%%%%%%%%%%%%%%%%%%%%%%%%%%%%%%%%%%%%%%%%%%%%%%%%%%%%%%%%%%%%%%%%%%%%%%%%
%%                            Introduction                               %%
%%%%%%%%%%%%%%%%%%%%%%%%%%%%%%%%%%%%%%%%%%%%%%%%%%%%%%%%%%%%%%%%%%%%%%%%%%%%%

\section{Introduction}

\begin{figure}[h!t]
\centering
\includegraphics[width=1.\textwidth]{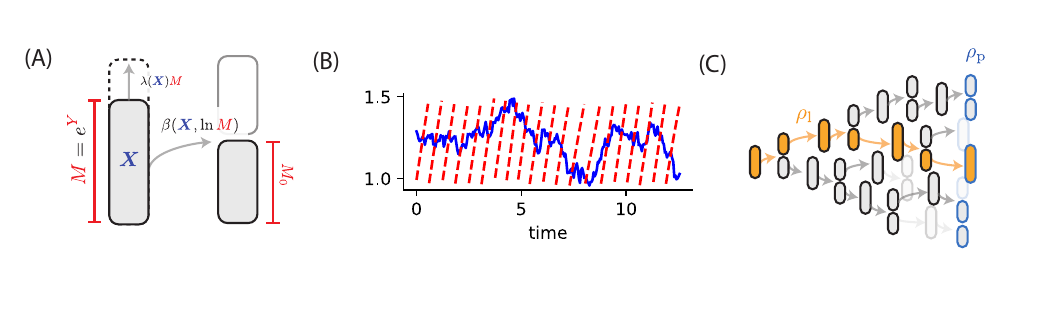}
\caption{(A) A diagram of the model. A cell has size ($Y$) and growth phenotypes ($X$). Division occurs at a rate that depends on both. We have omitted the explicit dependence on the initial size in this figure for simplicity. (B) A simulation of our model in the case when $X$ is an OU process, showing the evolution of $X$ (blue) and size $Y$ (red). (C) A diagram of a growing population showing the distinction between the lineage and population distributions.}\label{fig:1}
\end{figure}

In modern biology single-cell data are abundant, owing largely to microfluidic devices such as the \emph{mother machine} \cite{wang2010robust}, which enable precise measurements of gene expression and morphology within single-cell lineages. Traditionally, phenotype measurements (e.g., cell size, gene expression levels) were predominantly conducted in growing populations. Consequently, understanding how single-cell growth and division dynamics shape phenotype distributions in large populations has become essential. Specifically, in growing populations (Figure~\ref{fig:1}C), selection biases samples toward faster-growing cells, causing systematic differences between lineage-level and population-level distributions. Understanding the relationship between these distributions is thus critical, as it (1) reveals how populations respond to selective pressures and (2) allows inference of selection strength by comparing single-cell with population-level data
\cite{levien2021non,kussell2005bacterial,lewontin1969population}. 

There is a vast literature on these questions~\cite{kiviet2014stochasticity, lambert_quantifying_2015, hashimoto_noise-driven_2016, jafarpour2018bridging, lin2020single, nozoe2020,yamauchi2022unified, genthon_2022_jrsi, jafarpour2023evolutionary, hein2024competition, jia2021, pirjol2017phenomenology, jafarpour2019cell, barber_modeling_2021, kar2021, genthon2023, genthon2025noisy, elgamel2023, jia2021_prx, kohram2021, thomas2017single, Thomas2017making}, which can be traced back to foundational work by Euler -- which is summarized in the historical survey \cite{bacaer2011short}. In the mathematics literature, work on these topics falls under the umbrella of structured population models. Sharpe and Lotka were the first to study branching models with age dependence \cite{sharpe1911problem}. Powell later explicitly connected these models with experimental measurements of microbial population growth \cite{powell1956growth}. Many others have explored mathematical questions, such as existence and uniqueness, primarily using tools from semigroup theory \cite{engel2000one, Webb1983, Webb1985}. The importance of size regulation -- the process by which feedback from size to division dynamics maintains stable size distributions \cite{taheri2015cell}  -- was only recently appreciated due to work by Lin and Amir \cite{lin2017effects}. This has motivated considerable efforts in the biophysics community~\cite{nozoe2017inferring,yamauchi2022unified,jafarpour2019cell,genthon2025noisy,wang2023cell,qian2020counting}. These studies mostly focus separately on the dynamics of a fluctuating phenotype or on the mechanisms of cell-size regulation. 

In this paper, we study a model of size-regulated growth in which the per unit biomass growth rate, denoted $\lambda$, is determined by internal variables $\bX \in \reals^d$ (e.g., expression levels) that evolve independently of cell size between divisions, and may be affected at division.  We further assume that $\bX$ may affect the division rate, denoted $\beta$; see Figure~\ref{fig:1} A and B. Within this model, we explore (1) the structure of the stationary distribution of phenotypes in both the lineage and population ensemble, and (2) representations of population statistics in terms of lineage statistics, which are based on the Feynman--Kac formula. Previous work on Feynman--Kac representations has focused on models lacking the size variable \cite{marguet_2019_uniform}, but a basic fact is that the division rate $\beta$ cannot be independent of size. This is most easily seen in a model where $\lambda$ is fixed (see Section~\ref{sec:sizeonly}). In the absence of any feedback from size to the division rate, the log size will undergo a random walk, and therefore the variance in sizes will diverge \cite{amir2014cell}. Meanwhile, the dependence of $\lambda$ and $\beta$ on $\bX$ generates a potentially complex correlation structure between size and the phenotype $\bX$ within a population. This makes the problem of determining the distribution of $\bX$ difficult, because it requires marginalization over the size distribution. 

Interestingly, recent work \cite{hein2024asymptotic} has shown that in certain cases the size and growth rate are asymptotically decoupled, which means that they are independent variables in a snapshot of the population or a collection of lineages.  This phenomenon was also observed in \cite{tuanase2008regulatory}, in which the authors study the fitness effects of gene expression noise. When the growth rate and cell size decouple, we can study the growth rate distribution in a simpler model without a size variable. This has many practical advantages, such as making inference procedures robust to the model of cell size and making simulations more straightforward.  To show that size and growth decouple, in \cite{hein2024asymptotic}, the authors derive a lineage representation of population statistics. This has the form of a tilted expectation, which, as we will discuss, is closely related to the Feynman--Kac formula. 

In the model studied in \cite{hein2024asymptotic}, division is prescribed by a random threshold that may depend on the previous cell sizes. 
Our motivation was to understand more generally the relationship between the size and phenotype distributions. This is achieved by building upon the adder-sizer model in references~\cite{xia2020pde, auger2008structured} by combining it with the fluctuating growth rate model studied in \cite{hein2024asymptotic,levien2020interplay,tuanase2008regulatory,hein2024competition}. 

We specifically sought to answer the following questions: When do the size and phenotype asymptotically decouple in either the lineage or population distributions? When decoupling occurs, it is straightforward to represent the population statistics as a tilted expectation, which does not involve the size phenotype. This motivates another question:  What is the interpretation of this tilted expectation when decoupling does not occur? This leads us to a generalized Feynman--Kac relationship, which relates lineage statistics to a mass-weighted phenotype distribution.

\subsection{Organization of this paper}
In section~\ref{sec:model}, we introduce a general model for the dynamics of size-regulated single-cell growth, and we provide PDE descriptions of the population and lineage distributions. In section~\ref{sec:result}, we present our results about this model without much technical detail, while section~\ref{sec:examples} is devoted to some examples and numerical simulations.   In section~\ref{sec:timechange}, we show how the population distribution can be related to the lineage distribution via the Feynman--Kac formula when the size and growth phenotype decouple. This involves using a time change to ``tilt'', or importance sample, expectations of lineage observables to obtain population observables. In section~\ref{sec:mass_weighted} we provide a more general interpretation of the tilted expectation formula derived from the Feynman--Kac representation and connect our model to the discrete Feynman--Kac formula. In the more general case where size and growth phenotypes do not decouple, this can be understood as a size-biased average. Finally, we present the technical details and proof of our results in section~\ref{sec:proof}. For ease of reference, table~\ref{tab:notation} summarizes the notation used throughout, and table~\ref{tab:decoupling-cheatsheet} compiles the main results.

\begingroup
\setlength{\tabcolsep}{4pt}
\begin{table}[t]
\caption{Table of symbols and notations.}
\label{tab:notation}
\footnotesize\begin{tabular}{@{}ll}
\br
Symbol & Meaning \\
\mr
$\bX(t)\in\reals^d$ & Growth phenotype; forward operator $\mathcal L$; density $\nu(\bx,t)$\\
$\bY(t)=(y_c,y_b)$ & Size phenotype: current log-size $y_c$, birth log-size $y_b$ \\
$\lambda(\bx)$ & Biomass growth rate (single cell); $dM/dt=\lambda\big(\bX(t)\big)M$ \\
$\beta(\bx,\by)$ & Division rate (hazard); under Condition~\ref{eq:beta_assump}, $\beta=\lambda(\bx)\,\varphi(\by)$ \\
$h(\bx,y_c|\bx',y_c')$ & Post‑division kernel; under Condition~\ref{eq:h_assump}, $h=r(\bx|\bx')\,u(y_c|y_c')$ \\
$r(\bx|\bx')$ & Post‑division conditional growth phenotype density; under 
    Cond.~\hyperref[dec:SD]{SD}, $r\! =\! \delta(\bx\!-\!\bx')$ \\
$u(y_c|y_c')$ & Post‑division conditional log-size density; for 
    symmetric division, $u\! =\! \delta(y_c'\!-\!\ln2\!-\!y_c)$ \\
$\rho_{\ell},\,\rho_{\rm p}$ & Lineage and population joint densities; under decoupling, $\rho(\bx,\by)\!=\!\nu(\bx)w(\by)$\\
$\nu_{\ell},\,\nu_{\rm p}$ & Lineage and population growth‑phenotype densities\\
$w_{\ell},\,w_{\rm p}$ & Lineage and population size‑phenotype densities\\
$\E_{\ell}$, $\E_{\rm p}$ & expectations with respect to $\rho_{\ell}$ and $\rho_{\rm p}$ \\
$\Lambda$ & Malthusian growth rate (population) \\
$K(t)$ & Division count along a lineage (section~\ref{sec:tree}) \\
\br
\end{tabular}
\end{table}
\endgroup

%%%%%%%%%%%%%%%%%%%%%%%%%%%%%%%%%%%%%%%%%%%%%%%%%%%%%%%%%%%%%%%%%%%%%%%%%%%%%
%%                     Theoretical Framework                           %%
%%%%%%%%%%%%%%%%%%%%%%%%%%%%%%%%%%%%%%%%%%%%%%%%%%%%%%%%%%%%%%%%%%%%%%%%%%%%%

\section{Size and phenotype structured population model}\label{sec:model}

Here, we describe a class of single cell growth models in which a cell's phenotype at any instant can be decomposed into \emph{growth phenotype} $\bX(t)\in\mathbb{R}^d$ and \emph{size phenotype} $\bY(t) = (Y_c(t),Y_b(t)) \in \reals^2$. The growth phenotype, which could represent the concentration of growth-limiting cellular components such as ribosomes or proteins involved in antibiotic resistance, evolves independently of size between cell divisions according to some Markov process as described below. The size phenotype consists of the normalized (log) current cell size and size at the most recent division. Specifically, if $M_b(t)$ is the size at birth of a cell at time $t$, we set  $\bY(t) = (Y_c(t), Y_b(t))$ where $Y_c = \ln (M(t)/\E[M_b])$ and $Y_b(t) = \ln (M_b(t)/\E[M_b])$, and $\E[M_b]$ is the mean steady-state birth size.
{Meanwhile, cells accumulate size (e.g. mass) $M(t) = e^{Y_c(t)}\E[M_b]$ exponentially at a rate $\lambda(\bX(t))$; i.e.,
\begin{equation}\label{eq:Liouville}
Y_c'(t) = \lambda(\bX(t)),
\end{equation}and divide at a rate $\beta(\bX(t),\bY(t))$ yielding two newborn cells whose phenotypes are sampled from a distribution $h(\bx,y_c|\bx',y_c)$. }

 Note that the initial size is expressed as a function of time because it changes during cell division but remains constant within a cell cycle.  The motivation for keeping track of both the current size and size at birth is that many cells exhibit size-regulation mechanisms in which the growth increment between divisions depends on the initial size. To accommodate such regulatory strategies within our modeling framework, it is necessary to include both the current cell size and initial size. 

\subsection{Growth phenotype dynamics}
The dynamics of $\bX$ within the cell-cycle are characterized by a generator $\mcL$, which is a linear operator acting on a dense domain of functions on $\reals^d$. $\mcL$ defines a  size-independent $\tilde{\bX}(t)$ whose forward equation (known as the \emph{Fokker-Planck equation} for a diffusion process) is
\begin{equation}\label{eq:nufp}
\frac{\partial}{\partial t}\nu(\bx,t) = \mcL\nu(\bx,t)
\end{equation}
In the mathematics literature, one usually begins by defining the backward generator acting on test functions. In our notation this operator is $\mcL^*$, defined by
\begin{equation}
\mcL^*f(\bx) =\lim_{h \to 0}\frac{1}{h}\E[f(\tilde{\bX}(t+h))-f(\tilde{\bX}(t))|\tilde{\bX}(t) = \bx].
\end{equation}
Recall that solutions to the backward equation
\begin{equation}\label{eq:gfp}
\frac{\partial}{\partial t}g^f(\bx,t) = \mcL^* g^f(\bx,t),
\end{equation}
propagate expectations; that is, $g^f(\bx,t) = \E[f(\tilde{\bX}(t))|\tilde{\bX}(0) = \bx]$ with $g^f(\bx,0) = f(\bx)$. 
We use the convention that $\mcL$ (without the adjoint notation) propagates the probability density because this will appear more often in our presentation. As explained later, we will allow the possibility that $\mcL=0$, in which case $\tilde{\bX}(t) = \tilde{\bX}(0)$.

\subsection{Size and division dynamics}
We now discuss the joint process $(\bX(t),\bY(t))_{t \ge 0}$ which is built upon the size-independent process $\tilde{\bX}(t)$ as follows. 
For a cell born at time $t=0$ with growth phenotype $\bX(0) = \bx$, the growth phenotype distribution at time $t$ before division is obtained by propagating equation~\eqref{eq:nufp} with initial data $\nu = \delta_{\bx}$. The division time $\tau$ is determined by a division rate function $\beta$, defined as the per unit time probability to divide given that the cell has not yet divided:
\begin{equation}
\beta(\bx,\by)dt = \bbP\left(\tau \in [t,t+dt)|(\bX(t),\bY(t)) = (\bx,\by), \{\tau>t\} \right),
\end{equation}
After division, the distribution of newborn growth and size phenotypes is sampled from a conditional density
\begin{align}
\begin{split}
 &h(\bx,y_c|\bx',y_c')\prod_{i=1}^ddx_i dy_c\\
&= \bbP\left(\{\bX(t_k)  \in [\bx,\bx + d \bx)\} \cap \{Y_c(t_k)  \in [y_c,y_c + d y_c)\}|(\bX(t_k^-),Y_c(t_k^-)) = (\bx',y_c')\right)
\end{split}
\end{align}
where $t_k$ is the $k$th division event, $(\bX(t_k^-),Y_c(t_k^-))$ are the growth and size phenotypes immediately before division, and $[\bx,\bx + d \bx)$ is shorthand for the rectangle $[x_1,x_1 + dx_1)\times \cdots  \times [x_d,x_d + dx_d)$.

\noindent In summary, the model is defined by a quadruple $(\mcL,\lambda,\beta,h)$ containing:
\begin{itemize}
\item A linear operator $\mcL$ that is the generator of a continuous time Markov process (including the trivial case $\mcL =0$).
\item Growth and division rate functions $\lambda: \reals^d \to \reals_{> 0}$ and $\beta:\reals^d \times \reals^2 \to \reals$.   $\lambda$ is bounded below by a constant, while $\beta$ is an increasing function of the added size $Y_c-Y_b$ and vanishes as $Y_c \to Y_b$.
\item A transition kernel $h:(\reals^d \times \reals) \times (\reals^d \times \reals) \to \reals_{\ge 0}$ gives the phenotypes of a cell following a division event. $h(\cdot,\cdot|\bx,y_c)$ is a probability density that respects the conservation of mass at division-- that is, the sum of the two newborn daughter cell sizes must equal the mother cell size immediately before division. Said another way, if $y_c'$ is the log of the size of the mother cell before division and $y_c$ is the size of a daughter cell, then $e^{y_c'} - e^{y_c}$ (the mass of the other daughter cell) is equal in distribution to $e^{y_c}$. 
\end{itemize}
We note that without the growth phenotype, our model reduces to the adder-sizer model studied in references~\cite{xia2020pde,gabriel2018steady}. An additional discussion of the relationship to other models will be provided in section~\ref{sec:discuss}.

\subsection{Lineage distribution}
The lineage distribution is obtained when we consider the distribution of phenotypes over a sequence of cells, where at each division, one of the daughter cells is discarded. 
The evolution of the joint density $\rho_{\ell}(\bx,\by)$ of growth and size phenotypes, $(\bX_t,\bY_t) = (\bx,\by)$ along a lineage obeys the PDE
\begin{align}\label{eq:rho_pde}
\begin{split}
\partial_t \rho_{\ell}(\bx,\by,t) &= \mcA\rho_{\ell}(\bx,\by,t)
\end{split}
\end{align}
where 
\begin{equation}
 \mcA p(\bx,\by) = \mcL p(\bx,\by)- \lambda(\bx)\partial_{y_c} p (\bx,\by)- \beta(\bx,\by)p(\bx,\by)
\end{equation}
The advection term comes from equation \eqref{eq:Liouville}.

{The boundary condition at $y_c = y_b$ is obtained by examining the flux of probability mass in the $y_c$ direction at zero added size. Newborn cells have $y_c = y_b$, and because cells do not divide at the instant of birth, there is no loss of mass through this boundary. The only inflow comes from division events in the interior with $y_c' > y_b$ while the deterministic increase of added size generates an advective flux through the boundary equal to
\begin{equation}
\lambda(\bx)\,\rho_{\mathrm l}^{b}(\bx,y_b,t),
\end{equation}
where $\rho_{\mathrm l}^{b}(\bx,y_b,t)$ denotes the trace of the density at $y_c=y_b$. A mother cell with phenotype $(\bx',y_c',y_b')$ divides at rate $\beta(\bx',y_c',y_b')$ and produces a newborn daughter with phenotype $(\bx,y_b)$ with density $h(\bx,y_b \mid \bx',y_c')$. Integrating over all possible mother states gives the total rate at which newborns of type $(\bx,y_b)$ are created. Equating this influx with the outgoing advective flux yields
\begin{align}\label{eq:rholboundary}
\begin{split}
\lambda(\bx)\,\rho_{\mathrm l}^{b}(\bx,y_b,t)
= \int_{y_b}^{\infty}\!\int_{-\infty}^{y_c'} \!\int_{\mathbb{R}^d}
h(\bx,y_b \mid \bx',y_c')\,\beta(\bx',y_c',y_b')\,\rho_{\mathrm l}(\bx',y_c',y_b',t)\,d\bx'\,dy_b'\,dy_c'.
\end{split}
\end{align}}
Throughout, the superscript $b$ is used to denote the distribution of newborn cells.
% The boundary conditions are derived from examining the flux in the $y_c$ direction across the boundary at zero added size ($y_c = y_b$).  Cells do not divide at the instant of birth, so there is no loss of probability mass through this boundary. Instead, the boundary is fed solely by division events occurring in the interior with $y_c' > y_b$. The deterministic increase of added size induces an advective flux in the $y_c$--direction equal to
% \begin{equation}
% \lambda(x)\,\rho_{\mathrm l}^{b}(x,y_b,t),
% \end{equation}
% where $\rho_{\mathrm l}^{b}(x,y_b,t)$ denotes the trace of the density at the boundary.  On the other hand, a mother cell with phenotype $(x',y_c',y_b')$ divides at rate $\beta(x',y_c')$ and produces a newborn daughter with phenotype $(x,y_b)$ with probability density $h(x,y_b \mid x',y_c',y_b')$. The total rate at which newborn daughters of type $(x,y_b)$ are created by divisions is therefore given by  Equating this production rate with the advective flux yields 
% %In summary, assuming $\beta(\bx,\by)$ does not diverge at $y_c=y_b$ (cells do not divide at the instant of division), we have
% \begin{align}\label{eq:rholboundary}
% \begin{split}
% &\lambda(\bx)\rho_{\ell}^b(\bx,y_b,t) = \int_{y_b}^{\infty} \int_{-\infty}^{y_c'} \int_{\reals^d}h(\bx,y_b|\bx',y_c')\beta(\bx',y_c',y_b')\rho_{\ell}(\bx',y_c',y_b',t)d\bx'dy_b'dy_c'
% \end{split}
% \end{align}
% where
% \begin{equation}
% \rho_{\ell}^b(\bx,y_b,t) = \rho_{\ell}(\bx,(y_b,y_b),t).
% \end{equation} 
% Throughout this paper, we will use the superscript $b$ to denote the distribution over newborn cells. 

We are interested in the steady-state distributions
\begin{equation}
\rho_{{\ell},\infty}(\bx,\by) = \lim_{t \to \infty}\rho_{\ell}(\bx,\by,t)
\end{equation}
We remove the subscript $_\infty$ when it is clear that we are referring to the stationary density.  The calculation of $\rho_{\ell}$ is greatly simplified if the PDE equation~\eqref{eq:rho_pde} can be solved using separation of variables, which means that the stationary solution of the form $\rho_{\ell}(\bx,\by) = \nu(\bx)w(\by)$. In this case the $\bx$-factor $\nu(\bx)$ lies in the kernel of $\mcL$. Intuitively, we expect such a factorization to exist when $\beta$ and $h$ factorize, but as we explain below, this is not sufficient.

\subsection{Population distribution}
In a population in which cells undergo binary fission, we study the number density 
\begin{equation}
n(\bx,\by,t)d\bx d \by= \E[\#\{\text{Cells with $(\bX(t),\bY(t)) \in [\bx,\bx + d \bx) \times [\by,\by + d \by)$} \}].
\end{equation}
The expectation here is taken over different realizations of the population tree, which we call the \emph{population ensemble}. 
The number density is un-normalized, but it has a very similar structure to $\rho_{\ell}$. 
This is because, as long as $y_c>y_b$, the number of cells in the rectangle $[\bx,\bx + d \bx) \times [\by,\by + d \by)$ is not influenced by the division events. Meanwhile, the flux of cells in and out of this region due to the evolution of $\bX_t$, the accumulation of biomass, and the division events are all unaffected by the fact that we are dealing with a growing population \cite{xia2020pde}. Therefore $n(\bx,\by,t)$ obeys equation~\eqref{eq:rho_pde}.  The growth only manifests at the boundary where, instead of producing one cell, a division event produces two. As a result, the boundary conditions for $n$ are 
\begin{align}\label{eq:rholboundaryn}
\begin{split}
&\lambda(\bx)n^b(\bx,y_b,t) =2 \int_{y_b}^{\infty} \int_{-\infty}^{y_c'} \int_{\reals^d}h(\bx,y_b|\bx',y_c')\beta(\bx',y_c',y_b')n(\bx',y_c',y_b',t)d\bx'dy_b'dy_c'.
\end{split}
\end{align}

We take for granted that $n(\bx,\by,t)$ exists and exhibits balanced exponential growth, meaning
\begin{equation}\label{eq:nsep}
n(\bx,\by,t) \sim e^{\Lambda t}\rho_{\rm p}(\bx,\by). 
\end{equation}
Previous studies have rigorously derived conditions under which structured population models exhibit balanced exponential growth -- specifically, see references~\cite{marguet_2019_uniform,marguet_2019_lln,Webb1983}. 
The normalized density
\begin{equation}
\rho_{\rm p}(\bx,\by,t) = \frac{n(\bx,\by,t)}{\int \int n(\bx,\by,t)d\bx d\by},
\end{equation}
will be referred to as the population distribution.
The density of population phenotypes $\rho_{\rm p}(\bx,\by)$ can also be interpreted as the density of phenotypes in a chemostat, where cells are diluted at a rate $\Lambda$ \cite{levien2020interplay}. 
By plugging  equation~\eqref{eq:nsep} into equation~\eqref{eq:rho_pde} and equation~\eqref{eq:rholboundaryn}, obtain an eigenvalue problem for $\Lambda$. Just as with the lineage distribution, it is much easier to solve this problem when $\rho_{\rm p}$ can be factored. As we will see, when this is the case, the eigenvalue problem for the $\bx$ component is given by an eigenvalue problem for a \emph{tilted} version of the generator, which is closely related to the Feynman--Kac formula.

\section{Summary of results}\label{sec:result}

\subsection{Decoupling}

We will state our main result concerning decoupling. \textit{In words:} under biologically reasonable factorization and independence conditions, the long‑time joint phenotype distribution splits into the product of a size‑phenotype distribution and a growth‑phenotype distribution,
\begin{equation}\label{eq:decoupling}
\rho(\bx,\by,t) \to \rho(\bx,\by) = \nu(\bx)w(\by),\quad t\to \infty,
\end{equation}
where $\rho$ is either the lineage or population distribution. {As with $\rho$ will use the subscripts ${\rm p}$ and ${\ell}$ on $\nu$ and $w$ to indicate the population and lineage densities. }
The precise statement of this result is given in Theorem \ref{thm:main} below. 
We impose two structural conditions, which will be referenced throughout. 
\begin{assumplist}
  \item[(\mainassumplabel{eq:beta_assump})] $\beta(\bx,\by) = \lambda(\bx)\varphi(\by)$
  \item[(\mainassumplabel{eq:h_assump})] $h(\bx,y_b|\bx',y_c') = r(\bx|\bx')u(y_b|y_c')$
\end{assumplist}
{Condition \ref{eq:beta_assump} is equivalent to assuming that the division hazard per unit of (log-)size accumulation depends only on the size variables: since  $dY_c/dt=\lambda(X(t))$, the ``hazard per unit added log-size'' is $\beta/\lambda=\phi$. Thus, fluctuations in $\lambda$ modulate the speed of progression through the size-controlled division process rather than introducing an independent  ``chronological age'' clock. This is consistent with standard size-control modeling (adder/sizer-type regulation) \cite{jafarpour2019cell,amir2014cell} and it allows the growth process to be transformed to a process with constant growth rate via a random time-change $(T(t)=\int_0^t \lambda(X(s)),ds)$ (see Section \ref{sec:timechange}). }

{Condition  \ref{eq:h_assump}  corresponds to independent inheritance noise in the internal phenotype versus size partitioning noise, which may be reasonable in some settings but not universal. Section \ref{sec:mass_weighted} discusses how we can still obtain lineage representations of the population distribution when this condition fails.}

Note that compared to Condition \ref{eq:h_assump}, Condition \ref{eq:beta_assump} is much stronger; not only must $\beta$ factor, but the $\bx$ contribution must be exactly the growth rate.

Together, Conditions \ref{eq:beta_assump} and \ref{eq:h_assump} imply that the integral on the right-hand side of the boundary condition (equation~\eqref{eq:rholboundary}) can be factorized into a product of functions that depend separately on $\bx$ and $\by$, provided that $\rho(\bx,\by)$ is of the separable form $\rho(\bx,\by) = \nu(\bx) w(\by)$. Both conditions are necessary for this factorization because $\beta$ appears in the boundary conditions. However, even when both conditions hold, $\rho(\bx,\by)$ does not necessarily separate. This is because cell-cycle duration is determined by size. Thus, the time over which $\bX$ evolves between divisions depends on size, and consequently, the distribution of $\bX$ at division does too. Only when the distribution of $\bX$ is invariant under the boundary conditions can we expect to have some form of decoupling. We distinguish between strong and weak forms:

\begin{itemize}
  \item \textbf{Strong Decoupling (SD):} \phantomsection\label{dec:SD}
the boundary conditions leave $\bX$ unchanged ($r(\bx|\bx') = \delta(\bx -\bx')$) and $\mcL$ generates a stationary ergodic Markov process.  In this case  $\nu_{\rm p}$ obeys
 \begin{equation}\label{eq:eval}
\mcL\nu_{\rm p} + \lambda \nu_{\rm p} = \Lambda \nu_{\rm p}.
\end{equation}
  \item \textbf{Weak Decoupling (WD):} \phantomsection\label{dec:WD}
  The iteration
\begin{equation}\label{eq:Riter}
    \nu_{k+1} = R_{\lambda}\nu_k
\end{equation}
converges, where
\begin{equation}\label{eq:Rdef}
(R_{\lambda}\nu)(\bx) = \int r(\bx|\bx')\frac{\lambda(\bx')}{\lambda(\bx)}\nu(\bx')d\bx'
\end{equation}
and $\nu = \lim_{k \to \infty}\nu_k \in {\rm ker}\mcL$. {The most biologically relevant case is when $\mcL=0$, in which case this model becomes the random growth rate model discussed in Section \ref{sec:modelRG}.}
\end{itemize}

In the case of \hyperref[dec:SD]{SD}, equation~\eqref{eq:decoupling} holds for both $\rho_{\ell}$ and $\rho_{\rm p}$. The size distributions $w_{\ell}$ and $w_{\rm p}$ obey the same equations as the size-only model, while $\nu_{\ell} \in {\rm ker}(\mcL)$ and the population growth rate is determined by the eigenvalue problem, equation~\eqref{eq:eval}.
This is, in fact, the equation for the population density and asymptotic growth rate we would arrive at if we began with a model where there was no size regulation and cells divided at a rate $\lambda$. While it is a strong assumption that $\bX$ is unaffected by division events, recent analysis of high-resolution biomass data from cancer cells has shown this to be a good model of their growth rate fluctuations \cite{levien2025stochasticity}.  {Note that the case of a constant growth rate $\lambda(\bx) = \lambda_0$ is also covered by \hyperref[dec:SD]{SD}, since Condition  \ref{eq:beta_assump} automatically holds here. }

For \hyperref[dec:WD]{WD},  equation~\eqref{eq:decoupling} holds only for the lineage distribution. When $\bX$ is not continuous across division, we need the evolution of its probability density to be preserved in order to achieve some form of decoupling. However, since this evolution differs between the population and lineage distributions by a dilution term, we cannot have both distributions decouple in the same model (except for the trivial case of constant $\lambda$).

Note that the operator $R_{\lambda}$ is not the stochastic operator, $R_1$, corresponding to the integral kernel $r(\bx|\bx')$; however, it is similar to $R_1$ in the sense that $R_{\lambda} = \lambda^{-1}R_{1}\lambda$ (here $\lambda$ is interpreted as the multiplication operator that multiplies densities on $L^1(\reals^d)$ by $\lambda(\bx)$).
The existence of a stable, positive fixed point to the iteration equation~\eqref{eq:Riter} is guaranteed when $R_1$ is the generator of a discrete-time, irreducible, ergodic Markov chain. Moreover, if $\pi$ is probability density satisfying $R_1\pi = \pi$, the normalized fixed point of $\nu$ satisfying $R_{\lambda}\nu = \nu$ satisfies $\nu(\bx) = \pi(\bx)/\lambda(\bx) /\E_{\pi}[1/\lambda(\bx)]$. This mapping between densities is known as the Esscher Transform of $\pi$ (see \cite{elliott2005option}) and is usually written in terms of an energy function $E(\bx) = \ln \lambda(\bx)$. This transformation is also related to the Feynman--Kac formula, which interestingly emerges separately from the population statistics, as we describe below. 

{Finally, we note that a third case, which we refer to as Weak Decoupling in the population (WDp), may be considered. Here, decoupling occurs only in the population. However, because this does not have any natural biological meaning, we reserve a discussion to the technical details in Section \ref{sec:proof}.}

\subsection{Lineage representation of population statistics}
In addition to the decoupling results, we describe the relationship between lineage and population observables via path integral representations. These representations are also known as the ``Many-to-one'' formula, as they relate the population process (the many-body theory in physics terminology) to the lineage process \cite{marguet_2019_uniform,marguet_2019_lln}. 
For strong decoupling, we can obtain expectations with respect to the population distribution by averaging, yielding
\begin{equation}\label{eq:fk1}
\E_{\rm p}[f(\bX(t))] \sim \frac{\E_{\ell}^{\rm path}[f(\bX(t))e^{\int_0^t \lambda(\bX(s))ds}]}{\E_{\ell}^{\rm path}[e^{\int_0^t \lambda(\bX(s))ds}]}.
\end{equation}
where $\E_{\ell}^{\rm path}$ is used to represent averages with respect to the lineage sample path distribution on $[0,t)$ (rather than over $\nu(\bx,t)$).
Additionally, we will show that when division is symmetric, but the internal variable and size phenotype do not necessarily decouple, the right-hand side of equation~\eqref{eq:fk1} has an interpretation as an average with respect to a mass-weighted distribution (section~\ref{sec:mass_weighted}).

\section{Examples}\label{sec:examples}

\subsection{Size-only process}\label{sec:sizeonly}

{A special case of the model described above is the size-only process obtained by setting $\bX = \bx_0$ (by letting $\mcL=0$ and assuming \hyperref[dec:SD]{SD}) and hence $\lambda(\bx_0) = \lambda_0$. In this case, the phenotype of a cell is defined only by the size variable. }
The time-homogeneous solution to equation~\eqref{eq:rho_pde}, $\rho(\bx,\by)$ factors out into a growth distribution $\nu(\bx)$ and a size distribution $w(\by)$. The lineage density of the size variable $w_{\ell}(\by) = w_{\ell}(y_c,y_b)$ is given by 
\begin{equation}\label{eq:pdesizeonly}
\partial_{y_c} w_{\ell}(y_c,y_b) =  - \varphi(y_c,y_b)w_{\ell}(y_c,y_b).
\end{equation}
The full proof of this relationship is given in section~\ref{sec:proof}, but one can easily verify that it follows from equation~\eqref{eq:rho_pde} when growth is constant.

The size distribution of non-newborn cells can now be solved as 
\begin{equation}
\label{eq:sizeonly_lineagesol}
    w_{\ell}(y_c,y_b) = e^{-\int_{y_b}^{y_c}\varphi(z,y_b)dz} w_{\ell}(y_b,y_b).
\end{equation}
We introduce the marginal density of birth sizes
\begin{equation}
w_{\ell}^b(y) = Z^{-1}w_{\ell}(y,y),
\end{equation}
where the normalization constant $Z=\int_{-\infty}^\infty w_{\ell}(y,y) dy$ ensures this is a density in $y$. Using the boundary conditions given in equation~\eqref{eq:rholboundary}, we find that $w^b_{\ell}$ satisfies the integral equation 
\begin{equation}\label{eq:omegalsizeonly}
w_{\ell}^b(y) = \int_{-\infty}^{\infty}\int_{y_b'}^{\infty}u(y|y_c')f(y_c'|y_b')w_{\ell}^b(y_b')\, dy_c'dy_b',
\end{equation}
where 
\begin{equation}\label{eq:f}
f(y_c|y_b) = \varphi(y_c,y_b)e^{-\int_{y_b}^{y_c}\varphi(z,y_b)dz}
\end{equation}
is the density of the final log-fold change in size, $y_c$, conditioned on the initial cell-size $y_b$.

To obtain similar equations for the population birth size distribution, we need to start with the exponential growth ansatz for the size number density $n_{\rm p}(\by,t) \sim w_{\rm p}(\by)e^{\Lambda t}$ where $\Lambda$ is the asymptotic exponential growth rate. To find the population growth rate $\Lambda$, we follow \cite{lin2017effects} and look at the dynamics of the total mass of the cells in the population:
 \begin{align}
 \frac{d}{dt}\bar{m} &= \sum_{i=1}^N\frac{d}{dt}M_i(t) = \sum_{i=1}^N\lambda_i M_i(t) = \lambda_0 \bar{m}
 \end{align}
 where $\bar{m}$ is the total mass of the population, $M_i(t)$ is the mass of the $i$th cell, and $N$ is the total number of cells. The population growth rate is therefore given by $\Lambda = \lambda_0$  and is independent of the size-related parameters $\beta$ and $h$. Plugging $w_{\rm p}(\by)e^{\lambda_0 t}$ into  equation~\eqref{eq:rho_pde} and solving for the time-homogeneous solution yields 
 \begin{equation}\label{eq:sizeonly_popsol}
w_{\rm p}(y_c,y_b)  = w_{\rm p}(y_b,y_b)e^{-\int_{y_b}^{y_c}[\varphi(z,y_b) + 1]dz }.
\end{equation}
We note that the $+1$ that appears in the exponent comes from the fact that, as the population grows, the flux of newborn cells leads to the dilution of cells at any given size. This leads to the population version of equation~\eqref{eq:omegalsizeonly}: 
\begin{equation}\label{eq:omegapsizeonly}
w_{\rm p}^b(y) =2\int_{-\infty}^{\infty}\int_{-\infty}^{y_c'}u(y|y_c')f(y_c'|y_b')e^{-(y_c'-y_b')}w_{\rm p}^b(y_b')dy_b'dy_c'.
\end{equation}
In the special case of symmetric division, where $u(y_b|y_c) = \delta(y_b-y_c+\ln(2))$, one can check that for any solution $w_{\ell}(y_c,y_b)$ to the lineage size distribution equations~\eqref{eq:pdesizeonly} and \eqref{eq:omegalsizeonly}, the lineage distribution divided by cell mass $w_{\ell}(y_c,y_b)e^{-y_c}$ must satisfy the population distribution equations~\eqref{eq:sizeonly_popsol} and \eqref{eq:omegapsizeonly}. We therefore have the proportionality $w_{\rm p}(y_c,y_b) \propto w_{\ell}(y_c,y_b)e^{-y_c}$. In \ref{sec:apx_size_distribution}, we discuss the cell size distributions marginalized over birth size and show how they reduce to known forms that appear in other literature.

Notice that in order to obtain similar expressions to those in this section for the newborn size distribution in the general case, we need to marginalize over the sample path distribution of $\bX$. This is the fundamental problem we encounter in models with a fluctuating growth phenotype. The idea underlying our decoupling results is that when Condition \ref{eq:beta_assump} holds, cells have no intrinsic notion of age other than their size, allowing us to bypass the calculation of this path integral (see section~\ref{sec:timechange}). 

As an example, consider
\begin{equation}\label{eq:varphi_gauss}
\varphi(y_c,y_b) = \frac{1}{\sigma_Y}\frac{\phi\left(\frac{y_c-y_b(1-\alpha) - \ln(2)}{\sigma_Y}\right)}{1-\Phi\left(\frac{y_c-y_b(1-\alpha) - \ln(2)}{\sigma_Y}\right)}
\end{equation}
where $\phi$ and $\Phi$ are, respectively, the pdf and cdf of standard Gaussian distributions. The distribution of the final size conditioned on the initial size is approximately ${\rm Normal}(y_b(1-\alpha)+\ln(2),\sigma_Y)$, provided $\sigma_Y \ll 1$. This is the commonly used linear regression model for size dynamics \cite{hein2024asymptotic,cadart2018size,amir2014cell}.  The approximation comes from the fact that $f(y_c|y_b)$ is supported on $y_c\ge y_b$, whereas in the exactly Gaussian model $y_c$ may be less than $y_b$. We include a brief discussion of this autoregressive model and the biological significance of $\alpha$ in \ref{sec:apx_ar1}. Note that if we take $u(\cdot|y_c')$ to be Gaussian with mean $y_c'$ and variance $\sigma_d^2$, then we can absorb the variance introduced by division into the parameter $\sigma_Y^2$, thus we can set $u(y|y_c') = \delta_{y_c'-\ln(2)}$, without loss of generality. As long as $\alpha \in (0,2)$, the size distribution will remain stable.

\begin{figure}[h!t]
\centering
\includegraphics[width=0.9\textwidth]{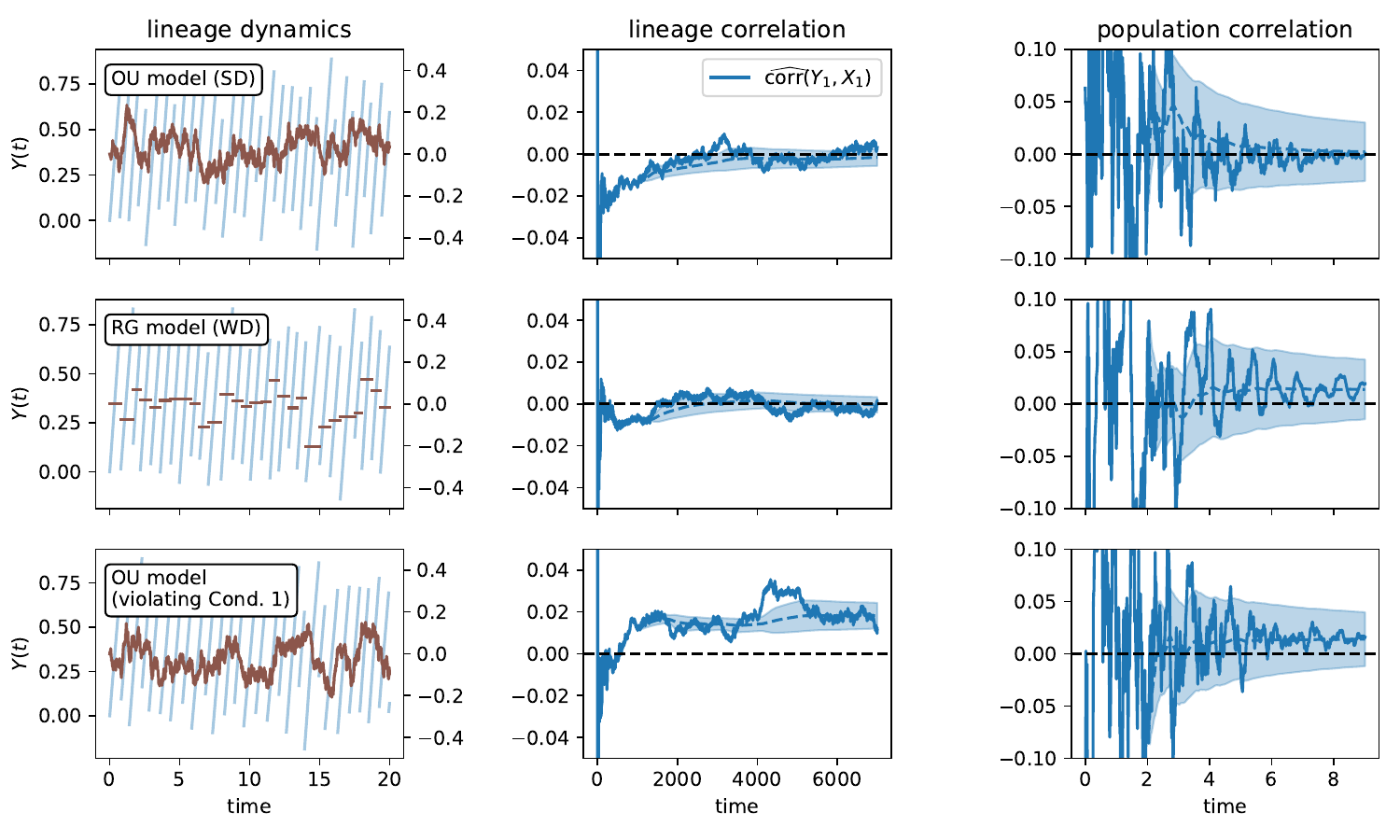}
\caption{(left) Simulations of three models illustrating examples of (top to bottom) (\hyperref[dec:SD]{SD}), (\hyperref[dec:WD]{WD}), and no decoupling. For all models $d=1$ and the growth rate function is $\lambda = 1 + x$. For (\hyperref[dec:SD]{SD}) we have simulated an OU process for the $\bX$ dynamics with $\theta = 1$, $\sigma^2 =0.01$, $\alpha=1/2$, $\sigma_Y = 0.01$. For (\hyperref[dec:WD]{WD}) we have simulated a model with $\mcL = 0$ and $r(\bx|\bx')$ is Normal distribution with mean $0$ and standard deviation $\sigma_x = 0.1$. The last model combines the OU dynamics with this division kernel.  (middle) The convergence of the empirical correlation coefficients between size variables and $X$ in the lineage and (right) population distribution.  The dashed lines show the running average of the correlations and the shaded area shows $\pm$ one standard deviation.  }\label{fig:2}
\end{figure}

\subsection{OU and RG Growth rate dynamics}\label{sec:modelRG}
As an example where the growth rate fluctuates, consider an Ornstein-Uhlenbeck (OU) process for the growth dynamics within the cell-cycle, which has previously been studied in references~\cite{tuanase2008regulatory,hein2024asymptotic,levien2021non,touchette2018introduction,du2023dynamical,buisson2022dynamical,levien2025stochasticity}. For simplicity we take $d=1$ (the multivariate version of this process is discussed in \ref{sec:apx_ou}). The process $X(t)$ is given by 
\begin{equation}\label{eq:OU_sde}
    dX(t) = -\theta X(t) dt + \sigma dW(t),
\end{equation}
where $\sigma \in \reals_{\ge0}$, $\theta \in \reals_{>0}$, and $W(t)$ is a standard Brownian motion, with the backward generator
\begin{equation}\label{eq:Lou}
\mcL^* = -x\theta \partial_x + \frac{\sigma^2}{2}\partial_x^2.
\end{equation}
The stationary density (which is the normalized Perron eigenvector of $\mcL$) of the size-independent process is a ${\rm Normal}(0,\sigma^2/(2\theta))$. 

We consider this model division rate $\beta = \lambda \varphi$, where $\lambda(x) = \lambda_0 + x$ and $\varphi$ is given by equation~\eqref{eq:varphi_gauss}. Although $\lambda$ is not strictly positive, 
we work in the small noise regime where the negative growth rates are exponentially unlikely; therefore, this can be considered an approximation to $(\lambda_0 + x)1_{x>-\lambda_0}$. 
As was shown in \cite{hein2024asymptotic}, if $r(x|x') = \delta(x-x')$ (which together with equation \eqref{eq:Lou} implies \hyperref[dec:SD]{SD}), $X(t)$ and $Y(t)$ are independent in both the lineage and population ensembles.  Note that technically, because $\lambda$ is not positive, we can obtain non-physical solutions where the growth rate becomes negative; however, but these have a negligible effect in the small noise regime. 
Figure~\ref{fig:2} (top row) illustrates this numerically. Here, we have computed the empirical correlation coefficients $\widehat{{\rm corr}}_T(X,Y_c)$. These show a clear convergence to zero as we looked over longer time windows (lineage dynamics) and larger population sizes (population dynamics).  

Within this model, we can also obtain an analytical formula for the population growth rate: $\Lambda = \lambda_0 + \E_p[x] = \lambda_0 + \sigma^2/(2\theta^2)$.
As claimed in section~\ref{sec:result} (see equation~\eqref{eq:eval}), this is indeed the largest eigenvalue of $\mcL + \lambda$. It could also be computed from equation~\eqref{eq:fk1}, which is a Gaussian integral since the integrated observable $\int_0^t\lambda(X(s))ds$ is a Gaussian.

%\subsection{Random growth rate model}\label{sec:modelRG}
We contrast the OU model with the so-called Random Growth rate (RG) Model where $\mcL =0$ and $r(\cdot|X')$ is taken to be a ${\rm Normal}(0,\sigma_X)$ density.  Simulations of this model are also shown in figure~\ref{fig:2}.  The RG falls into the case of weak decoupling (\hyperref[dec:WD]{WD}) if we use the same model for $\beta$ as in the previous subsection. Consistent with our results described above, the internal variable and size phenotype decouple in the lineage, but not in the population ensemble.Finally, we simulated the OU model but violated condition $1$ by setting $\beta = \lambda_0\varphi$, so there is not $\bx$ dependence on the division rate. As can be seen in the bottom row of figure~\ref{fig:2}, the correlations between $X$ and $Y_c$ persist as $t \to \infty$ both in the lineage and population.

%%%%%%%%%%%%%%%%%%%%%%%%%%%%%%%%%%%%%%%%%%%%%%%%%%%%%%%%%%%%%%%%%%%%%%%%%%%%%
%%             Eigenvalue and Path Integral Formulations                %%
%%%%%%%%%%%%%%%%%%%%%%%%%%%%%%%%%%%%%%%%%%%%%%%%%%%%%%%%%%%%%%%%%%%%%%%%%%%%%

\section{Strong decoupling and Feynman--Kac formula}\label{sec:timechange}

In this section, we discuss the eigenvalue problem in equation~\eqref{eq:eval} from the perspective of the Feynman--Kac formula. We do not go into the technical details of the Feynman--Kac theory, which can be found in \cite{moral2004feynman}. Applications to structured population models are discussed in \cite{marguet_2019_uniform,bertoin_2019_jfa}.

\subsection{Time-changed process and Feynman--Kac formula}
Under Condition \ref{eq:beta_assump}, the size dynamics can be transformed, via a random time-change, to a process $\widetilde{\bY}(t)$ with the same size distribution but a constant growth rate. Hence $\widetilde{\bY}(t)$ is exactly the size-only dynamics of section~\ref{sec:sizeonly}.

Recall that divisions occur at a rate $\varphi(\widetilde{\bY}(t))$ upon which $\widetilde{\bY}(t)$ is resampled according to the density $u(\by|\widetilde{\bY}(t))$. 
We define the stochastic time $T(t)$ by (see \cite{hein2024asymptotic})
\begin{equation}
T(t) = \int_0^t \lambda(\bX(s))ds, \qquad \frac{dT}{dt}=\lambda(\bX(t)). 
\end{equation}
The marginal size process generated by the model described in section~\ref{sec:sizeonly} is then given by 
\begin{equation}\label{eq:Ychanged}
\bY(t) = \widetilde{\bY}(T(t)). 
\end{equation}
Assuming $\lambda$ is positive, $T(t)$ is invertible, and we can write $ \widetilde{\bY}(s) =\bY(T^{-1}(s))$
This is because each time increment in the stochastic time variable corresponds to an increment $\lambda(\bX_t)dt$ in the original time variable. 

\begin{figure}[h!t]
\centering
\includegraphics[width=0.9\textwidth]{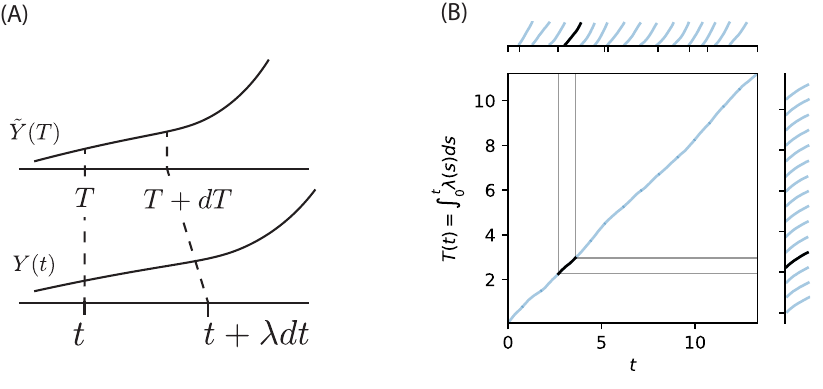}
\caption{(A) The time change converts the growth phenotype independent process $\tilde{\bY}(T)$ to $\bY(t)$. (B) The top panel shows the original size dynamics, while the right panel shows the same process when the cell size is plotted as a function of $T(t)$ rather than $t$.  }\label{fig:timechange}
\end{figure}

To derive equation~\eqref{eq:Ychanged}, consider the probability $\widetilde{\bY}(t)$ dividing (meaning it is resampled from $u$) in an interval $[T,T+dT)$, which, by the definition of $\varphi$, is
\begin{equation}
  \bbP\left(\text{$\widetilde{\bY}(T)$ divides in $[T,T+dT)$}\Big|\widetilde{\bY}(T)\right) = \varphi(\widetilde{\bY}(T))dT. 
\end{equation}
Hence, the probability for $\bY(t)$ to divide in $[t,t+dt)$ is  $\lambda(\bX(t))\varphi(\bY(t))dt$.
% \begin{align}
% %\begin{split}
%  & \bbP\left(\text{$\bY(t)$ divides in $[t,t+dt)$}\Big| \bY(t),\bX(t)\right)
%  % &=  \bbP\left(\text{$\widetilde{\bY}(T)$ divides in $[T(t),T(t) + T'(t)dt)$}\Big|\widetilde{\bY}(T),\bX(t)\right)\\
%   = 
% %  \end{split}
% \end{align} 
This transformation procedure is shown in figure~\ref{fig:timechange} (A).  In figure~\ref{fig:timechange} (B), we have shown trajectories of a process with an OU growth rate.  Note that the time change can be performed even when $\bX(t)$ depends on $\bY(t)$, because by the definition of $T(t)$, $T'(t)|\bX(t)$ is always deterministically equal to $\lambda(\bX(t))$. 

When (\hyperref[dec:SD]{SD}) holds, the population distribution may be obtained via expectations involving $e^{T(t)}$. 
Recall that in this case $r(\bx|\bx') =\delta(\bx-\bx')$ and $\mcL$ generate a stationary ergodic Markov process.  Let
\begin{equation}\label{eq:gdef}
g_{\rm p}^f(\bx,t) = \E_{\ell}^{\rm path}\left[ f(\bX(t))e^{T(t)}\Big|\bX(0) = \bx\right]. 
\end{equation}
We use the subscript ${\rm p}$ in anticipation of the fact that $g_{\rm p}^f$ will represent population expectations, in contrast to the expectations appearing in equation~\eqref{eq:gfp}.
Importantly, the expectation is over the lineage path distribution, since $T(t)$ depends on the entire trajectory of $\bX(t)$ from $0$ to $t$. 
According to the Feynman--Kac (FK) formula, $g_{\rm p}^f$ is the solution to a linear evolution equation \cite{touchette2018introduction}
\begin{equation}\label{eq:gzx}
\partial_t g_{\rm p}^f(\bx,t)  = \hat{\mcL}_{\lambda}^*\,g_{\rm p}^f(\bx,t),
\end{equation} 
with \emph{tilted generator}
\begin{equation}\label{eq:hatL}
\hat{\mcL}_{\lambda} = \mcL+ \lambda,
\end{equation}
and $g_{\rm p}^f(\bx,0) = f(\bx)$. Equation~\eqref{eq:gzx} can be derived for SDEs using It\^{o}'s lemma (see \cite{oksendal2013stochastic}), but it can also be derived for a more general process by looking directly at the forward variation of $g_{\rm p}^f$ and performing a Taylor expansion, similar to the textbook derivations of the Kolmogorov backward equations \cite{gardiner1985handbook}.

It is known that (see \cite{moral2004feynman}) the operator $\hat{\mcL}_\lambda$ has a dominant eigenvalue $\Lambda$ corresponding to a strictly positive, normalized right eigenfunction $\nu_{\rm p}$ (the population density); that is,  
\begin{equation}
\hat{\mcL}_{\lambda}\nu_{\rm p} = \Lambda\nu_{\rm p}, 
\qquad
\langle 1,\nu_{\rm p}\rangle = 1.
\end{equation}
Under mild assumptions on $\mcL$ (positivity and irreducibility), $\Lambda$ is algebraically simple and strictly dominates the rest of the spectrum, thus any solution $g_{\rm p}^f$ to equation~\eqref{eq:gzx} has the large-time asymptotics
\begin{equation}
g_{\rm p}^f(\bx,t) \;=\; e^{\Lambda t}\,\langle f,\nu_{\rm p}\rangle  + o(e^{\Lambda t}), \quad t\to\infty,
\end{equation}
where the coefficient $\langle f,\nu_{\rm p}\rangle$ is the projection of the initial data $f$ onto the right eigenspace. In particular, for $f\equiv 1$,
\begin{equation}
g_{\rm p}^1(\bx,t) \;=\; e^{\Lambda t} + o(e^{\Lambda t}).
\end{equation}
Taking the ratio for large $t$,
\begin{equation}\label{eq:Epop_M}
\lim_{t \to \infty}\frac{g_{\rm p}^f(\bx,t)}{g_{\rm p}^1(\bx,t)} =
\frac{\langle f,\nu_{\rm p}\rangle}{\langle 1,\nu_{\rm p}\rangle} 
= \int f(\bx)\,\nu_{\rm p}(\bx)\,d\bx.
\end{equation}
Notice that 
\begin{equation}
\frac{g_{\rm p}^f(\bx,t)}{g_{\rm p}^1(\bx,t)}= \frac{\E_{\ell}^{\rm path}[f(\bX(t))e^{T(t)}]}{\E_{\ell}^{\rm path}[e^{T(t)}]}
\end{equation}
The ratio appears in equation~\eqref{eq:fk1}. Therefore, as claimed, the population-level expectations can be obtained as the $t\to\infty$ limit of weighted lineage expectations.

Equation~\eqref{eq:fk1}  can also be stated as a formula for the bias between the population and lineage distributions as 
\begin{equation}
     \E_{\rm p}[f(\bX(t))] -  \E_{\ell}[f(\bX(t))]  = {\rm Cov}_{\ell}^{\rm path}\left( f(\bX(t)), \frac{e^{T(t)}}{\mathbb{E}_{\ell}\left[e^{T(t)}\right]}\right).
\end{equation}
These relationships connect the quantities of interest in microbial population models to importance sampling, where one weights samples by an exponential factor \cite{siegmund1976importance} to sample more efficiently from a different distribution. In effect, exposing a population to selection is one procedure to ``importance sample'' the phenotypes.

\begin{figure}[h!t]
\centering
\includegraphics[width=.8\textwidth]{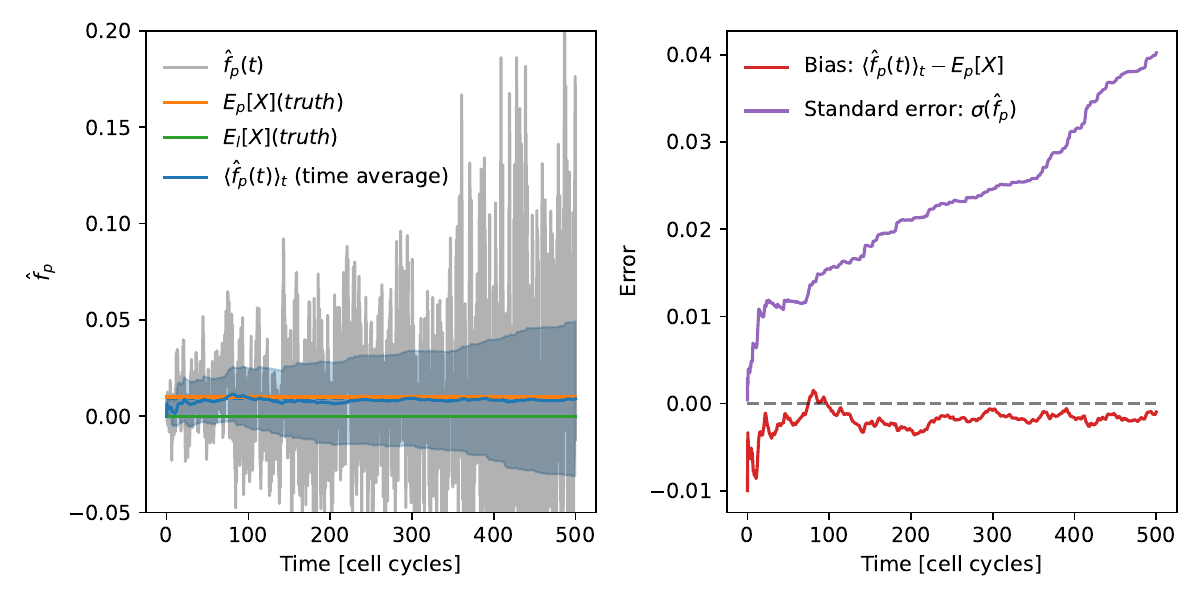}
\caption{{(left) Estimated phenotypic average $\hat{f}_p(t)$ compared to theoretical expectations $E_{\rm p}[X]$ and $\E_{\ell}[X]$, with cumulative time average shown in blue. (right) Error showing deviation of time-averaged estimate from true value and standard error. Results from $m= 100$ independent lineage simulations of OU with $T = 1000$ with $\sigma^2= 0.01$, $\theta =1$ and $\sigma_Y = 0.05$. }}\label{fig:fk_numerics}
\end{figure}

\subsection{Numerical example}
To numerically illustrate equation~\eqref{eq:fk1}, we obtain population expectations from lineage data using the OU process example from section~\ref{sec:examples}. In this case, we have an analytical formula for the population growth rate and population expectation of $\E_{p}[x]$. Equation~\eqref{eq:fk1} says that we can in principle retrieve this quantity from $m$ independent realization of the lineage process, $\{(\bX_i(s),\bY_i(s))_{0 \le s \le t}\}_{i=1,\dots,m}$. Note that the stochastic time $T_i(t)$ for each lineage can be obtained directly from the size measurements using
\begin{equation}\label{eq:T_disc_sum}
T(t) = Y_c(t)-Y_b(t) + \sum_{i=1}^{K(t)}(Y_c(t_i^-) -Y_b(t_i^-)),
\end{equation}
where $t_i^-$ is the time just before the $i$-th division along the lineage with $t_0=0$.

{For the $m$ lineages, we compute the estimate
\begin{equation}
\hat{f}_p(t)= \frac{\sum_{i=1}^mf(\bX_i(t))e^{T_i(t)}}{\sum_{i=1}^me^{T_i(t)}}.
\end{equation}
of $\E_{\rm p}[f(\bX(t))]$. In figure~\ref{fig:fk_numerics}, we have plotted this estimator as a function of time for $m= 500$ lineages. We also plot the time-average $\langle \hat{f}_p\rangle_t = \frac{1}{t}\int_0^t\hat{f}_p(s)ds$ and the running standard deviation. We observe that $\langle \hat{f}_p\rangle_t$ tends towards the true population average, although the variance of this estimator diverges. }

{To explain the behavior in Figure~\ref{fig:fk_numerics}, note that the Law of Large Numbers ensures that for fixed $t$, as $m \to \infty$, $\hat{f}_p(t) \to \mathbb{E}_\ell[f(X(t))\,e^{T(t)}]/\mathbb{E}_\ell[e^{T(t)}]$. The equivalence between this lineage-based quantity and the population expectation holds only asymptotically in time; thus, so the lineage must be run for a burn-in period. However, the variance of $e^{T_i(t)}$ grows exponentially in $t$, which implies that longer burn-in periods yield larger standard errors. Consequently, accurate estimation of the population expectation requires, in principle, exponentially many lineages. The exact scaling between time and the required number of lineages depends on the model and the observable. Nevertheless, in the present example, the estimator performs well even over time scales that can realistically be sampled in a mother-machine experiment.}

{A similar situation is considered in the context of growth rate estimation in references~\cite{levien2020large,grandpre2025extremal,rohwer2015convergence}, 
In particular, it is shown that there are two sources of error, a finite time bias and a sampling bias, the interplay of which leads to non-monotonic convergence in $t$.
These works focus on the estimation of the dominant growth rate $\Lambda = \lim_{t \to \infty}t^{-1}\ln\E_{\ell}[e^{T(t)}]$, rather than the expectations with respect to $\nu_{\rm p}$, but we expect $\hat{f}_{\rm p}(t)$ to be plagued by the same issues.}

\section{Symmetric division and the mass-weighted phenotype density of path expectations}\label{sec:mass_weighted}
This section provides a more general interpretation of the tilted expectation equation~\eqref{eq:gdef}, which holds when the size and growth phenotypes do not decouple, yet division is symmetric.The argument is based on previously developed lineage representations of population observables developed in \cite{nozoe2017inferring} and later used to relate independent lineage statistics to population growth rates in ~\cite{levien2020large,hein2024asymptotic,grandpre2025extremal}. We review this in the following subsection.

\subsection{The population tree argument and population distribution}
\label{sec:tree}

Consider a population starting from one cell at time $t=0$ and select a lineage from a tree by randomly choosing either daughter cell at time $t$. In the following, we will consider the distributions of the $N(t)$ lineages in this tree.

{For a given realization of the population tree, a lineage can be sampled in a forward manner by selecting a random cell at each division, which induces a distribution $\bbP_{f|{\rm tree}}$.  The notation $f|{\rm tree}$ is to remind us that we are conditioning on the tree. If the lineages are labeled $1,\dots,N(t)$, the probability of sampling a particular lineage with label $i$ is $2^{-K_i}$, where $K_i$ is the number of divisions along lineage $i$.}

{A lineage can alternatively be sampled in a backward manner by randomly selecting a lineage from the $N(t)$ cells at time $t$. The probability of sampling a label $i$ under $\bbP_b$ is $1/N(t)$. Performing a change of measure relates the expectations for any observable $f(\bX,\bY)$ via 
\begin{equation}
\E_{b|{\rm tree}}[f(\bX_i,\bY_i)] = \frac{1}{N(t)}\E_{f|{\rm tree}}[f(\bX_i,\bY_i)2^{K_i}]. 
\end{equation}
Note that because we are conditioning on the tree structure, we can move $N(t)$ outside of the expectation. }

{Note that averaging the expectations of $\bbP_{f}$ and $\bbP_b$ over trees yields asymptotically $\E_l$ and $\E_p$ respectively. Hence 
\begin{align}\label{eq:pbackrelate}
\E_p[f(\bX,\bY)] &\sim \E[\E_{b|{\rm tree}}[f(\bX_i,\bY_i)]]
= \E\left[\frac{1}{N(t)}\E_{f|{\rm tree}}[f(\bX_i,\bY_i)2^{K_i}]\right]
\end{align}
where the outer expectation on the right is over the tree distribution.
Note that if $f(\bX,\bY) = 1$, then 
\begin{equation}
\E[\E_{f|{\rm tree}}[2^{K_i}]/N(t)] =1. 
\end{equation}
Since $N(t) \sim e^{\Lambda t}$, 
this yields $\E_{f|{\rm tree}}[2^K_i] \sim Ce^{\Lambda t}$ where $C$ is a time-independent constant. 
The lineage expectation is related to the expectation on the right hand side of equation~\eqref{eq:pbackrelate} via
\begin{equation}
 \E\left[\frac{1}{N(t)}\E_{f|{\rm tree}}[f(\bX_i,\bY_i)2^{K_i}]\right]
 \sim e^{\Lambda t}\E_l^{\rm path}[f(\bX(t),\bY(t))2^{K(t)}]
\end{equation}
Combining these asymptotics we can relate expectations of the population observable to path expectations:
\begin{equation}
\label{eq:Epop_N}
    \E_{\rm p}\left[ f(\bX(t),\bY(t))\right] \sim \frac{\E_{\ell}^{\rm path}\left[ f(\bX(t),\bY(t))2^{K(t)}\right]}{\E_{\ell}^{\rm path}\left[ 2^{K(t)}\right]}.
\end{equation}
This is a general formula that holds irrespective of the division and growth dynamics. For example, even if $\mcL$ depends on the size variable, we can still obtain population expectations in this way. }

\subsection{Connecting Feynman--Kac expectation to mass-weighted statistics}

We want to interpret equation~\eqref{eq:Epop_M} within the context of a model where decoupling does not occur.  In the case of symmetric division, $Y_b(t_i^-)=Y_c(t_{i-1}^-)-\ln(2)$. Replacing this relation in the sum given by equation~\eqref{eq:T_disc_sum},  canceling out all the intermediate $Y_c(t_i^-)$s, as well as using $Y_b(t) = Y_b(t_{K(t)}^-)$ simplifies equation~\eqref{eq:T_disc_sum} to
\begin{equation}
T(t) = Y_c(t) + \ln(2)K(t).
\end{equation}
Hence, the exponential weight $e^{T(t)}$ from equation~\eqref{eq:gdef} splits into a mass term $e^{Y_c(t)}$ for the current cell and a multiplicity term $2^K(t)$ for the number of cells present
\begin{equation}\label{eq:EleTEleK}
\E_{\ell}[f(\bX(t))e^{T(t)}] = \E_{\ell}[f(\bX(t))e^{Y_c(t)}2^{K(t)}].
\end{equation}
In particular, $\E_{\ell}[e^{Y_c(t)}2^{K(t)}]$ is the expected total biomass of the population at time $t$. 

{Using the number density, $n(\bx,\by,t) \sim e^{\Lambda t}\rho_{\rm p}(\bx,\by)$, we can introduce a mass-weighted population density of the phenotype $\bX$ defined by
\begin{equation}\label{eq:mass_dist}
m(\bx,t) = \int e^{y_c}n(\bx,\by,t) d\by.
\end{equation}
The left hand side of equation~\eqref{eq:Epop_N} is, by definition, an expectation with respect to $\rho_{\rm p} \sim e^{-\Lambda t}n(\bx,\by,t)$. Meanwhile, equation~\eqref{eq:EleTEleK} is of the form of the numerator on the right hand side of equation~\eqref{eq:Epop_N} when we make the replacement $f(\bX,\bY) \to f(\bX)e^{Y_c}$. It follows that
\begin{equation}
    \E_{\ell}[f(\bX(t))\,e^{Y_c(t)} 2^{K(t)}] \sim \int f(\bx)\, m(\bx,t)\, d\bx,
\end{equation}}
Thus, in general, tilted expectations of the form given in equation~\eqref{eq:gdef} can be understood as averages with respect to this size-weighted density, and the finite-time version of equation~\eqref{eq:Epop_M} can be interpreted as the population mass-weighted expected value of $f(\bX)$. 

Only when we have the decoupling condition, we can pull out the expected value of $e^{Y_c(t)}$ from both the numerator and the denominator of equation~\eqref{eq:Epop_M}, recovering the count-based population distribution given in equation~\eqref{eq:Epop_N}. Nevertheless, we can still interpret equation~\eqref{eq:Epop_M} as the mass-weighted expected value, even in the absence of the decoupling, and even in finite time.

\section{Operator splitting view and proof of results}\label{sec:proof}

\subsection{Setup}
Here, we provide a formulation of the model in terms of linear operators between certain function spaces, without assuming Conditions \ref{eq:beta_assump} and \ref{eq:h_assump}. We find this to be helpful in understanding the mathematical origin of decoupling, which is due to a type of operator splitting \cite{macnamara2017operator} whereby the operators describing growth and division factor when applied to the dominant eigenvector.
We begin by introducing some notation.
Let
\begin{align}
\bbY = \{(y_c,y_b) \in \reals^2:y_c\ge y_b\}, \quad \bbY_b =\{(y_c,y_b) \in \reals^2:y_c = y_b\} \subset \bbY.
\end{align}
 Representing the space of size phenotypes and the space of newborn cells (note $\bbY_b \equiv \reals$).
The linear operators will act on subspaces of the function spaces $\mcM_x = L^1(\reals^d)$, $\mcM_y = L^1(\bbY)$, $\mcM_y^b = L^1(\bbY_b)$, $\mcM =  L^1(\reals^d \times \bbY )$ and $\mcM^b =  L^1(\reals^d \times \bbY_b )$. These contain the relevant probability densities.

\subsubsection*{Lineage equations}
We will start with the lineage distribution and omit their subscript $_{\ell}$.
The boundary conditions in our model define a linear operator $B:\mcM \to \mcM^b$ which maps the joint density of growth phenotype, size, and initial size to the density at $y_c = y_b$.  $B$ acts on a function $\rho \in \mcM$ according to 
\begin{align}
\begin{split}
(B \rho)(\bx,y_b) &=\frac{1}{\lambda(\bx)} \int_{y_b}^{\infty} \int_{-\infty}^{y_c'} \int_{\reals^d}h(\bx,y_b|\bx',y_c')\\
&\quad\quad\quad\quad\quad\quad\times \beta(\bx',y_c',y_b')\rho(\bx',y_c',y_b')d\bx'dy_b'dy_c'
\end{split}
\end{align}
This is simply the right-hand side of equation~\eqref{eq:rholboundary} written as a linear operator and divided by $\lambda(\bx)$. We are using parentheses around $B\rho^b$ to make it clear that the input is a function of $\bx$ and the full size phenotype $\by$, while the output is evaluated at $(\bx,y_b)$. 

We next introduce an operation which takes functions in $\mcM^b$ and lifts them to $\mcM$. This operation is obtained from a $C_0$-semigroup $(\Pi(y))_{y \ge0}$ on $\mcM$, meaning  $\Pi(y)\rho_b$ is a continuous function from $\reals_{\ge 0}$ to $ \mcM^b$ for any $\rho_b \in \mcM^b$. This comes from solving the time-independent version of equation~\eqref{eq:rho_pde} (the PDE for $\rho_{\ell}$) and treating size as the time variable. More concretely, for $\rho^b \in \mcM^b$ we define $G:\mcM^b \to \mcM$ by
\begin{equation}
  (G\rho^b)(\bx,y,y_b) = (\Pi(y-y_b)\rho^b)(\bx,y_b)
\end{equation}
where the function obtained by the application of $\Pi(y)$ to a function $\rho^b \in \mcM^b$, $u(\bx,y,y_b) =  (\Pi(y)\rho^b)(\bx,y_b)$, is a solution to
\begin{equation}\label{eq:rho_vs_y1pde}
 \partial_{y}u(\bx,y,y_b) = \tilde{\mcA}u(\bx,\by),\quad \tilde{\mcA}  = \frac{1}{\lambda(\bx)}\mcL  - \frac{\beta(\bx,\by)}{\lambda(\bx)} 
\end{equation}
at ``time'' $y$ starting with initial data $u(\bx,0,y_b) = \rho^b(\bx,y_b)$.  The rationale for this definition is: In order to obtain the steady-state distribution $\rho$ evaluated at $\bx,y_c,y_b$ from $\rho^b$, we need to evolve equation~\eqref{eq:rho_vs_y1pde} for $y_c - y_b$ which is the (log) size we need to add to get from the boundary to size $y_c$.

The unique steady-state linear density satisfies the equation
\begin{equation}\label{eq:rhol_op}
\rho_{\ell} = G B\rho_{\ell}.
\end{equation}
Note that because equation \eqref{eq:rhol_op} is equivalent to the time-independent Kolmogorov equation for the lineage dynamics, it has a unique, normalizable solution.

\subsubsection*{Population equations}
The situation for the population distribution is similar. As explained in section~\ref{sec:model} the only difference is a dilution term in the growth dynamics and a factor of two in the boundary conditions. To handle the dilution factor, we introduce 
a $\Lambda$-dependent family of $C_0$-semigroups $(\hat{\Pi}_{\Lambda}(y))_{y \ge 0}$ which produce solutions of
\begin{equation}\label{eq:rho_vs_y1pde_pop}
 \partial_{y_c}u(\bx,y_c,y_b) = \tilde{\mcA}u(\bx,\by)
 - \frac{\Lambda}{\lambda(\bx)}u(\bx,\by).
\end{equation}
up to ``time'' (size added) $y$. Let
\begin{equation}
  (\hat{G}_{\Lambda}\rho_b)(\bx,y,y_b) = \hat{\Pi}_{\Lambda}(y-y_b)\rho^b(\bx,y_b).
\end{equation}
Similar to equation~\eqref{eq:rhol_op}, we have
\begin{equation}\label{eq:eval_general}
\rho_{\rm p} =2 \hat{G}_{\Lambda} B\rho_{\rm p}
\end{equation}
These equations are a type of nonlinear eigenvalue problem that must be solved for both the density and $\Lambda$.

%For the dominant eigenvalue $\Lambda$, the operators $2\hat{G}_{\Lambda}B\rho_{\rm p}$ and $2B\hat{G}_{\Lambda}\rho_{\rm p}^b$ are generally not stochastic operators, but equations~\eqref{eq:eval_general} still have unique normalizable solutions. 

\subsection{Model structure under conditions \ref{eq:beta_assump} and \ref{eq:h_assump}}

\subsubsection*{Lineage equations}
We now show how the operator viewpoint makes the decoupling criteria transparent, beginning with the lineage distribution. First, we factor the ‘project‑to‑boundary’ operator $B$; next, we factor the ‘grow‑from‑boundary’ semigroup $G$. The product of these pieces acts separately when applied to the dominant eigenfunction, yielding the results stated in Section~\ref{sec:result}.  Under Conditions \ref{eq:beta_assump} and \ref{eq:h_assump}, $B = R_{\lambda} \otimes U_{\varphi}$ where $R_{\lambda}: \mcM_x \to \mcM_x$ is given by equation~\eqref{eq:Rdef} and $U_{\varphi}:\mcM_y \to \mcM_y^b$ acts on functions according to
\begin{equation}
(U_{\varphi} w)(y_b) = \int u(y_b|y_c')\varphi(y_c',y_b')w(y_c',y_b')d\by'.
\end{equation}
The subscripts remind us that these operators depend, respectively, on $\lambda$ and $\varphi$, hence incorporating the two factors in the division rate function separately.

Meanwhile, Condition \ref{eq:beta_assump}
allows us to factor the matrix exponential solution of ~\eqref{eq:rho_vs_y1pde}. Specifically, $\tilde{\mcA}$ becomes 
\begin{equation}
\tilde{\mcA}=\frac{1}{\lambda(\bx)}\mcL - \varphi(\by).
\end{equation}
and the second term, a multiplication operator, only depends on $\by$. As a result, the two operators are commutative and the operator exponential can be split into the exponential of $\mcL$ and a multiplication by a function $\psi$ depending on the size variables: 
\begin{equation}
\Pi(y_c-y_b)\rho^b(\bx,y_b)= \psi(y_c,y_b)\,e^{\frac{y_c-y_b}{\lambda(\bx)}\mcL}\rho^b(\bx,y_b),
\end{equation}
with 
\begin{equation}
\psi(y_c,y_b) = e^{-\int_{0}^{y_c-y_b}\varphi(y_b+z,y_b)\,dz}.
\end{equation}
Note that $G$ does not factor into a product as $B$ does into operators that act separately on $\mcM_x^b$ and $\mcM_y^n$. 
The important property is that $G$ leaves ${\rm ker}(\mcL)$ fixed, because the action in the $\bx$ component is entirely from the operator exponential of $\lambda^{-1}\mcL$, whose kernel is the same as $\mcL$. The factor $\lambda^{-1}$ corresponds to the $\bX$ dependent time change described in section~\ref{sec:timechange}, which appears because we propagate the $\bX$ dynamics in the size, rather than the time variable. 

Finally, note that we can relate $G$ to the size-only process by introducing a multiplication operator $\Psi: \mcM_y^b \to \mcM_y$ defined by $(\Psi w_b)(y_c,y_b) = \psi(y_c,y_b) w_b(y_b)$.
The steady-state size distribution is a fixed point of $\Psi U$.

\subsubsection*{Population equations}
For the population distribution equations, we have 
\begin{equation}\label{eq:hatPi_split}
\hat{\Pi}_{\Lambda}(y_c-y_b)\rho^b(\bx,y_b)= \psi(y_c,y_b)\,e^{\frac{y_c-y_b}{\lambda(\bx)}\left(\mcL - \Lambda\right)}\rho^b(\bx,y_b),
\end{equation}
In order to connect this to the eigenvalue problem, equation~\eqref{eq:eval}, we introduce 
\begin{equation}
\hat{\psi}(y_c,y_b) = e^{-\int_{0}^{y_c-y_b}[\varphi(y_b+z,y_b)-1]\,dz}.
\end{equation}
so that equation~\eqref{eq:hatPi_split} becomes 
\begin{equation}
\hat{\Pi}_{\Lambda}(y_c-y_b)\rho^b(\bx,y_b)= \hat{\psi}(y_c,y_b)\,e^{\frac{y_c-y_b}{\lambda(\bx)}\left(\mcL - (\Lambda - \lambda(\bx)) \right)}\rho^b(\bx,y_b),
\end{equation}
Observe that $\nu \in {\rm ker}(\mcL -  (\Lambda - \lambda))$ if and only if $\nu$ solves equation~\eqref{eq:eval}; therefore, $\hat{G}_{\Lambda}$ leaves solutions to equation~\eqref{eq:eval} unchanged. 

Just as we found for the lineage equations, the steady-state population size distribution for the size-only process is a fixed point of the operator $2\hat{\Psi}U$ where $\hat{\Psi}$ multiplies function in $\mcM_y^b$ by $\hat{\psi}(y_c,y_b)$.

\subsection{Statement of proof of main result}
{We are now ready to prove the result in section~\ref{sec:result}. Our result includes an extension of the previously stated conditions, which will not be of biological interest but will help provide a more complete picture of the mathematical structure. 
\begin{itemize}
  \item \textbf{Weak Decoupling in population (WDp): }\label{dec:WDp}The iteration in equation \eqref{eq:Riter}
converges to the unique solution to the eigenvalue problem in equation \eqref{eq:eval}.
\end{itemize}
This form of decoupling requires fine-tuning of the growth phenotype division kernel to the within-cell-cycle dynamics. It should be noted that \hyperref[dec:WD]{WD} and \hyperref[dec:WDp]{WDp} are mutually exclusive unless the single–cell growth rate $\lambda(\bx)$ is constant. Indeed, suppose there exists a positive density $\nu$ such that $\rho_\ell(\bx,y)=\nu(\bx)w_\ell(y)$ and $\rho_{\rm p}(\bx,y)=\nu(\bx)w_p(y)$. Then $\nu$ lies both in $\ker {\mcL}$ (lineage PDE) and solves $(\mcL+\lambda)\nu=\Lambda\nu$ (population PDE), so subtracting gives $\lambda(\bx)\nu(\bx)=\Lambda\nu(\bx)$. If $\nu$ has full support, $\lambda(\bx)\equiv\Lambda$ is constant. Thus, for nonconstant $\lambda$, simultaneous decoupling of lineage and population requires strong decoupling (SD), where the boundary kernel leaves $\bX$ unchanged, $r(\bx\mid \bx')=\delta(\bx-\bx')$. In the trivial constant–$\lambda$ case, \hyperref[dec:WD]{WD} and \hyperref[dec:WDp]{WDp} can both hold without r being delta–distributed, but then $\bX$ does not affect growth and there is no selection bias between ensembles.}

\begin{theorem}\label{thm:main}
Suppose Conditions \ref{eq:beta_assump} and  \ref{eq:h_assump} hold; that is,  $(\mcL,\lambda,\beta,h) = (\mcL,\lambda,\lambda\varphi,ru)$. Additionally, assume the lineage and population distributions converge to unique time-invariant steady-states $\rho_{\ell}$ and $\rho_{\rm p}$. Then, we have the following: 
\begin{itemize}
    \item In the case of \hyperref[dec:SD]{SD}, decoupling (equation~\eqref{eq:decoupling}) holds for both the lineage and population distributions. 
    \item In the case of \hyperref[dec:WD]{WD} (resp. \hyperref[dec:WDp]{WDp}), equation~\eqref{eq:decoupling} holds for the lineage (resp. population) distribution.
\end{itemize} 

\end{theorem}

\subsection{Proof}  

{The Theorem is a straightforward consequence of the formalism developed above.  In particular, if the separable solutions do indeed solve equation~\eqref{eq:rhol_op} (resp. equation~\eqref{eq:eval_general}) for the lineage (resp. population) distributions, then by assumption these are the unique, stable steady-state phenotype distributions. }

\subsubsection*{Lineage distribution} 
{The separable solution, if it exits, is $\rho = \nu \cdot w$, where $\nu \in {\rm ker}(\mcL)$ and $w$ is the unique stationary densities of the size only process, i.e. $\Psi\circ U w = w$. 
In the case of \hyperref[dec:SD]{SD}, $R_{\lambda}=I$ and therefore applying $G$ to $\nu \cdot w$ gives 
\begin{equation}
GB(\nu \cdot w) =  G(\nu \cdot Uw) = \nu \cdot \Psi Uw = \nu \cdot w. 
\end{equation}
In the case of \hyperref[dec:WD]{WD}, $R_{\lambda} \ne I$, but $R_{\lambda}\nu =\nu$ by assumption and therefore
\begin{equation}
GB(\nu \cdot w) =  G(\nu \cdot Uw) = R_{\lambda}\nu \cdot \Psi Uw = \nu \cdot w. 
\end{equation}
Thus, in both \hyperref[dec:SD]{SD} and \hyperref[dec:WD]{WD}, $\rho = \nu \cdot w$ is a fixed point. }

\subsubsection*{Population distribution} 
The proof for the population cases are identical, with the exception that instead of $\nu \in {\rm ker}(\mcL)$ we have $\nu \in {\rm ker}(\mcL - (\Lambda - \lambda))$.

\begingroup
\setlength{\tabcolsep}{6pt}
\begin{table}[h!]
\caption{Decoupling regimes and population-lineage relations.}
\label{tab:decoupling-cheatsheet}
\footnotesize
\begin{tabular}{@{}P{2.1cm}P{4.1cm}P{3.5cm}P{4.6cm}@{}}
\br
Regime & Condition & What factorizes & How to compute \\
\mr
Strong decoupling (SD) & $\,\beta=\lambda\,\varphi$, $\,h=r\,u$ with $r(\bx|\bx')=\delta(\bx-\bx')$; $\bX$ stationary ergodic & Lineage \emph{and} population:\quad
$\rho_{\ell}(\bx,\by)=\nu_{\ell}(\bx)w_{\ell}(\by)$ 
$\rho_{\rm p}(\bx,\by)=\nu_{\rm p}(\bx)w_{\rm p}(\by)$&
$\nu_{\ell}\in\ker\mathcal L$;
$\nu_{\rm p}$ solves $(\mathcal L+\lambda)\nu_{\rm p}=\Lambda\nu_{\rm p}$; $w_{\ell},w_{\rm p}$ are the size‑only solutions (section~\ref{sec:sizeonly});
$\displaystyle \E_{\rm p}[f(\bX(t))]=
\frac{\E_{\ell}^{\rm path}\!\left[f(\bX(t))e^{\int_0^t\lambda(\bX(s))ds}\right]}
     {\E_{\ell}^{\rm path}\!\left[e^{\int_0^t\lambda(\bX(s))ds}\right]}$.\\
\mr
Weak decoupling in lineage (WD) & $\,\beta=\lambda\,\varphi$, $\,h=r\,u$; $R_\lambda$ iteration converges and its fixed point lies in $\ker\mathcal L$ & Lineage only:
$\rho_{\ell}(\bx,\by)=\nu_{\ell}(\bx)w_{\ell}(\by)$ &
$\nu_{\ell}$ is the fixed point of $\nu\mapsto R_\lambda\nu$ (equation~\eqref{eq:Rdef});
$w_{\ell}$: size‑only;
population generally \emph{does not} factorize.\\
\mr
Weak decoupling in population (WDp) & $\,\beta=\lambda\,\varphi$, $\,h=r\,u$; $R_\lambda$ iteration converges to the unique solution of $(\mathcal L+\lambda)\nu=\Lambda\nu$ & Population only:
$\rho_{\rm p}(\bx,\by)=\nu_{\rm p}(\bx)w_{\rm p}(\by)$ &
$\nu_{\rm p}$ from $(\mathcal L+\lambda)\nu_{\rm p}=\Lambda\nu_{\rm p}$;
$w_{\rm p}$: size‑only with dilution;
lineage generally \emph{does not} factorize.\\
\mr
Symmetric division, no decoupling & $u$ symmetric (two equal daughters), no factorization assumed & — &
For any $t$:
$\displaystyle \E_{\rm p}^{\rm mass}[f(\bX(t))]=
\frac{\E_{\ell}^{\rm path}\!\left[f(\bX(t))e^{\int_0^t\lambda(\bX(s))ds}\right]}
     {\E_{\ell}^{\rm path}\!\left[e^{\int_0^t\lambda(\bX(s))ds}\right]}$;
here $\E_{\rm p}^{\rm mass}$ is with respect to the mass‑weighted density $m(\bx,t)$ (equation~\eqref{eq:mass_dist}).\\
\mr 
General case & — & — & $\E_{\rm p}\left[ f(\bX(t))\right] = \frac{\E_{\ell}\left[ f(\bX(t))2^{K(t)}\right]}{\E_{\ell}\left[ 2^{K(t)}\right]}$\\
\br
\end{tabular}
\end{table}
\endgroup

%%%%%%%%%%%%%%%%%%%%%%%%%%%%%%%%%%%%%%%%%%%%%%%%
%%                   Discussion and Conclusion                         %%
%%%%%%%%%%%%%%%%%%%%%%%%%%%%%%%%%%%%%%%%%%%%%%%%%%%%%%%%%%%%%%%%%%%%%%%%%%%%%

\section{Discussion}\label{sec:discuss}
The dynamic interplay of growth (i.e., biomass accumulation) and the cell cycle is a distinctive feature of models of cell growth. Much has been done to understand how the details of these processes influence size homeostasis and population dynamics. Here, we generalized work on decoupling, lineage--population relations, and pathwise Feynman--Kac formulations~\cite{hein2024asymptotic, garcia2019linking, levien2020large, bressloff2017feynman, lin2017effects, lin2020single} to describe the structure of the phenotype distribution under different decoupling scenarios, as summarized in table~\ref{tab:decoupling-cheatsheet}. We worked within a general model, accommodating most existing models of size regulation and cell growth. Within this model, a cell's phenotype is separated into two distinct phenotypes: a growth phenotype, which determines the biomass accumulation rate, and a size phenotype, which contains information about a cell's absolute size and its newborn size.  We have demonstrated that, in the lineage distribution, even when growth is controlled by a phenotype perturbed at division events, the size distribution remains insensitive to the details of how this phenotype evolves. When the growth phenotype is ``blind'' to division events, the size distribution and growth phenotype distribution are independent in both distributions. 

The decoupling of the growth and size phenotypes emerges due to a type of operator splitting, where the propagation of phenotypes from the boundary of newborn cells and the division dynamics act separately on the steady-state distribution (even though these operations do not commute).
In the most general case, we have shown that a size-biased sampling yields a generalized Feynman--Kac representation of population expectations. Prior work has rigorously proved Feynman--Kac formula for structured population models (see reference~\cite{marguet_2019_uniform}), but not in the context of size-structured models with fluctuating growth rates. Practically, these formula allow us to use lineage-tracked data to obtain population statistics, and they motivate importance‑sampling approaches for estimating population expectations from lineage paths. 

Our results have potential applications and extensions. First, they could be used to predict population distributions in liquid culture based on lineage averages. It is often assumed that these two types of experiments are equivalent, but rigorous comparison requires an understanding of how to link single–cell observables to population dynamics. Our results lay out a framework for doing this under certain constraints.

A second direction for future work is to explore how the decoupling and approximate–decoupling regimes we have characterized could be used to accelerate Feynman–Kac–based simulation algorithms, such as the fixed–budget Gillespie scheme of \cite{chen2025fixed}, for example by using the decoupled or approximately decoupled dynamics as low–variance proposals or control variates to reduce the variance of Monte Carlo estimators in these methods.

We also believe that the operator formulation of the steady-state dynamics (resulting in equations~\eqref{eq:rhol_op} and \eqref{eq:eval_general}) is of interest outside the decoupling context. This could potentially be a starting point for constructing spectral representations of the population and lineage distributions, which contain information about the transient dynamics.

%%%%%%%%%%%%%%%%%%%%%%%%%%%%%%%%%%%%%%%%%%%%%%%%%%%%%%%%%%%%%%%%%%%%%%%%%%%%%
%%                            Acknowledgments                          %%
%%%%%%%%%%%%%%%%%%%%%%%%%%%%%%%%%%%%%%%%%%%%%%%%%%%%%%%%%%%%%%%%%%%%%%%%%%%%%

\subsection*{Acknowledgments}

We thank Tom Chou for helpful discussions and two anonymous referees for their detailed comments, which significantly improved the quality of this manuscript. We also thank the organizers of the 2025 Gordon Research Conference on Stochastic Physics in Biology, which helped facilitate progress on this paper.

\vspace{1cm}
\appendix 
\addtocontents{toc}{\protect\setlength{\cftsecnumwidth}{7em}}

\section{Cell size distribution}\label{sec:apx_size_distribution}

Another interesting distribution is simply the distribution of cell sizes randomly selected from a single lineage, which the marginal density of cell size $y_c$ is defined as
\begin{equation}
    w_{\ell}^1(y) = \int_{-\infty}^{y}  w_{\ell}(y,y_b)dy_b.
\end{equation}
This can be shown to be the difference between the cumulative distribution functions of birth size and division size
\begin{equation}\label{eq:wl1sol}
    w_{\ell}^1(y) = \int^y_{-\infty} dy_b  w^b_l(y_b) - \int_{-\infty}^y dy_d  w_l^d(y_d)
\end{equation}
where the division size density distribution is defined as
\begin{equation}\label{eq:omegaldivision}
     w_l^d(y) =\int_{-\infty}^{y}dy_b f(y|y_b) w_l^b(y_b)
\end{equation}
In the case of symmetric division, where $u(y_b|y_c)=\delta(y_b-y_c+\ln(2))$, it follows from equation~\eqref{eq:omegalsizeonly} and equation~\eqref{eq:omegaldivision}, that $ w_l^d(y) =  w_l^b(y-\ln(2))$. Consequently, the lineage size distribution can be simplified even further as
\begin{equation}\label{eq:wl1sol_symmetric}
    w_{\ell}^1(y) = \int_{y-\ln(2)}^ydy_b  w_l^b(y_b)= F_b(y) - F_b(y-\ln(2)),
\end{equation}
Where $F_b(y)$ is the cumulative birth size distribution along a lineage. Similarly, the marginal density of cell size in a population is defined as
\begin{equation}
    w_{\rm p}^1(y) = \int_{-\infty}^y w_{\rm p}(y,y_b)dy_b.
\end{equation}
For symmetric division, we argued that there is a proportionality $w_{\rm p}(y_c,y_b) \propto w_{\ell}(y_c,y_b)e^{-y_c}$, hence the population cell size distribution is
\begin{equation}
    w_{\rm p}^1(y) \propto e^{-y} \left[F_b(y) -F_b(y-\ln(2))\right],
\end{equation}
which matches the result from reference~\cite{hein2024asymptotic}.

\section{Connection to autoregressive models for size dynamics}\label{sec:apx_ar1}
Previous works have used a linear autoregressive model for the size dynamics wherein the value of $Y_c-y_b$ at cell division is conditionally Gaussian
\cite{hein2024asymptotic,cadart2018size}: 
\begin{equation}\label{eq:example_Y2dist}
Y_c(\tau)|Y_b \sim {\rm Normal}(\ln(2) + (1- \alpha) Y_b  ,\sigma_Y^2). 
\end{equation}
The constant $\alpha$ is exactly the regression coefficient of initial size on size-added, known as the cell-size control parameter. A special significance is ascribed to the values $\alpha=1/2$ and $\alpha=1$. Known as adder and sizer strategies, respectively, these represent models of regulation where the cell adds an approximately constant size increment between divisions (adder) or attempts to divide at a constant size (sizer). Cell size regulation strategies ranging from adder to sizer are widely adopted by a host of organisms across the tree of life \cite{sauls2016adder,rhind2021cell,ho2018modeling,amir2014cell,cadart2018size}.

\section{Formula for multivariate OU process}\label{sec:apx_ou}

Here, we give the formula for the dominant eigenfunction $\nu_{\rm p}$ of the tilted generator (that is, the solution to equation~\eqref{eq:eval}) for the multidimensional OU process case with quadratic growth rate function. These formula generalize those presented in section~\ref{sec:examples} for $d=1$ and a linear growth rate.  The formula comes from previous calculations of the large deviation rate function and the scaled cumulant generating function (SCGF) \cite{du2023dynamical,buisson2022dynamical,benitz1990large,bercu1997large,bryc1997large,gamboa1999functional}. These results are also known in the context of models with deterministic division \cite{tuanase2008regulatory}. We have shown they hold more generally due to the strong decoupling.

The generator of the process is 
\begin{equation}
\mcL^* = -\sum_{i=1}^d\sum_{j=1}^d\Theta_{i,j} x_i \partial_{x_j} +\sum_{j=1}^d \sum_{i=1}^dD_{i,j}
\partial_{x_j} 
\partial_{x_i} 
\end{equation}
where $\Theta$ and $D$ are $d \times d$ non-singular matrices. The growth rate function is
\begin{equation}\label{eq:growthrate_quad}
\lambda(\bx) = \lambda_0 + {\bf b}^T\bx + \bx^TA\bx.
 \end{equation}
where ${\bf b} \in \reals^d$ and $A \in \reals^{d \times d}$.
    
It is well known that the lineage density of $\bx$ in the decoupling regime (i.e., the kernel of $\mcL$) is a multivariate Gaussian with covariance matrix $\Sigma$ given from the Lyapunov equation
\begin{equation}
    \Sigma \Theta^T + \Theta \Sigma = 2D.
\end{equation}
The population distribution $\nu_p$ is Gaussian distributed with the covariance $\Sigma_{\rm p}$ and mean $\bar{\bx}$ are given by 
\begin{equation}
    2 A - \Theta^T\Sigma_{\rm p}^{-1} - \Sigma_{\rm p}^{-1} \Theta + 2\Sigma_{\rm p}^{-1} D\Sigma_{\rm p}^{-1}  =0,
\end{equation}
and
\begin{equation}
    \bar{\bx} =\Sigma \left( 2\Sigma^{-1}D-\Theta^T \right)^{-1}  {\bf b}.
\end{equation}
Note that in the special case where the growth rate is a linear function, $\lambda(\bx) = \lambda_0 + {\bf b}^T \bx$, the covariance matrix of the population phenotype reduces to the regular Lyapunov equation.

\section*{References}
\bibliographystyle{unsrt}

\bibliography{./pop_growth_decoupling.bib}

\end{document}